# Statistical Mechanical Approach to Human Language

Kosmas Kosmidis, Alkiviadis Kalampokis and Panos Argyrakis

Department of Physics, University of Thessaloniki, 54124 Thessaloniki, Greece.

**Abstract**

We use the formulation of equilibrium statistical mechanics in order to study some important characteristics of language. Using a simple expression for the Hamiltonian of a language system, which is directly implied by the Zipf law, we are able to explain several characteristic features of human language that seem completely unrelated, such as the universality of the Zipf exponent, the vocabulary size of children, the reduced communication abilities of people suffering from schizophrenia, etc. While several explanations are necessarily only qualitative at this stage, we have, nevertheless, been able to derive a formula for the vocabulary size of children as a function of age, which agrees rather well with experimental data.

# 1. Introduction

Human language has recently become a subject of interdisciplinary character. Linguistic studies have traditionally been qualitative rather than quantitative. Recently, some attempts based on evolutionary game theory [1] have been made in an effort to understand language evolution, which have yielded some noticeable results. Particularly, interesting considerations were made in studies of competition between languages using mathematical [2] and computational models [3-10].

In this paper we propose the assumption that human language can be described as a physical system within the framework of equilibrium statistical mechanics. Defining a Hamiltonian analogue that is associated with words, we are able to explain basic properties of spoken languages, such as the universality of the exponent of Zipf law [11], and to predict reasonably well the form of the curve for the vocabulary size versus age for young children. We, thus, demonstrate that statistical physics can provide an interesting formulation for the study of spoken languages and can unify aspects, such as the frequency distribution of words and the children's vocabulary learning rate, properties which at first glance seem completely different.

A rather remarkable feature, common to several languages is the so called Zipf law [1], which states that if we assign the value m=1 to the most frequent word of a language, m=2 to the second one etc. then the frequency of occurrence of a word with rank m is

$$C_m \sim m^{-a} \tag{1}$$

This law has been verified experimentally for several languages with the exponent $α$ value found to be universal and approximately equal to one. An alternative way, which is also used in the literature, to present Zipf law is to state that the proportion of words $p_f$ whose frequency is $f$ (taking values in the range 0 to 1) in a given sample text is modelled by a power function $p_f \sim f^{-\beta}$. The exponent $\beta$ is related to the exponent $\alpha$ in equation (1) with the equation $1/\alpha = \beta - 1$. Although it is not immediately evident, the frequency-rank Zipf plot is equivalent to a plot of the cumulative distribution of $p_f$ versus frequency $f$ [12, 12]. Ref [12] in particular, contains a detailed proof of the above statement.

Traditionally statistical mechanics does not deal with human language. It deals with physical systems, i.e. with collections of atoms, molecules or other elementary particles. According to statistical mechanics, when a system of particles is in equilibrium at constant temperature $T$, then it can be found in one of $N$ states. The probability that it is found at a given state $i$ with energy $E_i$ is proportional to $\mathrm{Exp}(-\frac{E_i}{k_B T})$, the "Boltzmann factor." The temperature $T$ is the "measure" of the interaction of the system with the environment.

## 2. The Basic Assumption of the Model

Suppose that an individual possesses a vocabulary of $N$ words. We treat the language department of the human brain as a physical system that can be found in one of $N$ states. Each state represents one word. There is a one-to-one mapping between these states (which are enumerated using integers up to $N$) and words in the individual's vocabulary. If the system is found in state i then the word associated with state i is pronounced. We denote as "temperature", $T$, a measure of the willingness (or ability) that the language speakers have in order to communicate. Common sense indicates that some words are more useful than others. The word "food" is essential and no organized group of people will go very far without it in their vocabulary. The word "heterogeneous" is probably not so useful since groups of people will probably survive without knowing it. Of course, usefulness is not only associated with meaning. For example, the word "and" is very useful because it is used to connect words. In this case, the usefulness originates from syntax rather than meaning.

Our *ansatz* is that the Hamiltonian of the system is $H(k) = \varepsilon\, lnk$, where $\varepsilon$ is a constant and $k$ is a measure of a words usefulness, i.e. we assign the value $k=1$ to the most useful words, $k=2$ to the second ones etc.

Following the basic assumption of statistical mechanics, the probability to find a word with usefulness $k$ is

$$p(k) = \frac{1}{Z} \mathrm{Exp}(\frac{-H(k)}{k_B T}) = \frac{1}{Z} k^{-\varepsilon/k_B T} \qquad (2)$$

where $Z$ is the partition function of the system [14]. Note, however, that equation (2) is not Zipf law. Zipf law connects the frequency of occurrence of words in a language with the rank of the word. Equation (2) relates the probability of occurrence of a word with its usefulness $k$. Although we expect words of low $k$ value to have a low rank $m$, and vice versa, there is no reason prohibiting two or more words from having the same value of $k$ while these words have different rank.

**3. Implications of the Model**

*a. Divergence from the Power Law at the Initial Part of a Zipf Plot*

It has been observed that while equation (1) is a straight line in log-log form, there are noticeable deviations in the early part of the line [15]. In Figure 1 we plot the word occurrence, $C_m$, versus the word rank, m, using experimental data that come from a corpus (large collection of texts) consisting of publications in several Greek Internet sites up to May 2001, collected by Prof. Franz Guenthner at the University of Munich.. It has been checked and used by T. Kyriakopoulou [16]. This corpus contains a total of about 2.6 x $10^7$ words, about 2 x $10^5$ different words and about 5.5 x $10^4$ different lemmas. According to the Zipf law, such a frequency–rank plot should be a straight line in log-log plot as the word frequency distribution is a power law. This is observed in Figure 1, where the straight line is the best fit. Moreover equation (2) implies that in the frequency–rank plot we will be able to identify groups of words having different rank but the same frequency. Indeed this can be seen in the figure and is the reason for the divergence from the straight line at the initial part of the curve at small *m* values. Groups of words having the same frequency exist also in the rest of the curve but are particularly visible in the beginning, as this part is amplified by the log- log plot.

*b. Universality of the Zipf Exponent*

Equation (2) additionally implies that the exponent of the Zipf law is related to the "temperature" T. Since the usefulness of a word *k* is expected to be more or less proportional to its rank then we expect that the Zipf exponent *α*:

$$\alpha \approx \frac{\varepsilon}{k_B T} \qquad (3)$$

We also expect the communication willingness of the writers to be statistically the same (on the average) in different groups of people, such as English, German, Greek writers, etc. The temperature *T* of each group will be the same and the exponent *α* is expected to have a universal value, as is indeed experimentally observed.

*c. Emergence of Syntactic Communication.*

The partition function of the system is calculated as follows: Suppose that on the average there are *b* words having the same usefulness. Then the partition function may be approximated[1] by the following integral:

---

[1] An exact calculation of the partition function is also possible using harmonic numbers. At this point however, it is not necessary, since our main objective is to get only a rough idea of how the model can actually be used.

$$Z \approx \int_{1}^{N/b} b\, k^{-\varepsilon/k_B T} dk = k_B T [N(\frac{b}{N})^{\varepsilon/k_B T} - b]/(k_B T - \varepsilon). \tag{4}$$

From the partition function we can now calculate all the properties of the system, such as the average energy:

$$E = -\frac{\partial \ln Z}{\partial \beta} = \varepsilon [\frac{k_B T}{k_B T - \varepsilon} + \ln(\frac{b}{N})(\frac{b}{b - N(\frac{b}{N})^{\varepsilon/k_B T}} - 1)] \tag{5}$$

Note the fact that $E$ becomes indeterminate when $k_B T = \varepsilon$. It is useful to examine the significance of this behavior. It is due to the fact that the integral in eq. (4) produces the given solution only when $\frac{\varepsilon}{k_B T} \neq 1$. Moreover, the partition function $Z$ remains finite while the number of word $N$ approaches infinity, only if $k_B T < \varepsilon$. Thus, if $k_B T < \varepsilon$ it is allowed to have a language with infinite number of words. In this case, there is no limitation for the effectiveness of the language as it can be used to describe an infinite collection of items, meanings, ideas etc. When $k_B T > \varepsilon$ the partition function diverges when $N$ approaches infinity. Thus, there is a limitation for the language effectiveness.

For species with low communication willingness (or ability) a language comprised of words only is sufficient to cover all possible communication needs since an infinite number of words is possible. On the other hand this is not true for species with high $T$. They have to find a way to overcome this difficulty. It seems that the way to do it is by developing a language based on syntactic communication, as syntax allows us to formulate an almost unlimited number of sentences combining a limited number of words. This result agrees with the observation that animal communication is non syntactic whilst human language is syntactic. The need for syntactic communication was also a result in the treatment of Nowak [17], derived in a completely different way, namely the assumption of finite learning time, and thus the existence of a maximum number of words in an individual lexicon.

*d. Evolution of the Vocabulary Size of Children.*

In order to be able to apply our model to data of language development in children vocabulary we will make the following assumptions: It is known that children language development is related to the speech they hear. At the age of 16 months the children vocabulary is rather small, but over the subsequent months it shows accelerating growth [18]. We will approach the vocabulary growth of children as a warming process. The temperature in this case is a measure of communication ability. We assume that this ability increases proportionally with time. This is, of course, the simplest choice possible. Thus, we set

$$T = \gamma t \qquad (6)$$

where γ is a proportionality constant which may be different for different individuals. It is a measure of an individual's learning ability. We assume that a word is known if the probability to appear is greater or equal to a threshold $p_c$. Estimation for the spoken vocabulary size, i.e. the number of different words that the child will use as it speaks, is possible as follows:

The usefulness $k$ of the rarest observable word is denoted by $k_{max}$ and is calculated by solving the equation $p(k_{max}) = p_c$.

Thus,

$$\frac{1}{Z} k_{max}^{-\varepsilon/k_B T} = p_c \qquad (7)$$

There is of course the obvious constrain that $k_{max}$ has values in the interval [1, $N/b$]

Thus, the vocabulary size[2] is

$$V_s(t) = b(k_{max} - 1) = b[-1 + (\frac{A}{p_c})^{\gamma k_B t/\varepsilon}], \qquad (8)$$

where $A$ is equal to the following expression:

$$A = \frac{-1}{b - N(\frac{b}{N})^{\varepsilon/(\gamma k_B t)}} + \frac{\varepsilon}{(b - N(\frac{b}{N})^{\varepsilon/(\gamma k_B t)})\gamma k_B t} \qquad (9)$$

We use the above equation in order to compare the model with experimental data taken from Ref 5. The resulting Figure 2 shows that, despite approximations, the model describes the experimental data reasonably well. For producing Figure 2 we have used the value $p_c$=0.0001 for the threshold[3] and the value $N$=6000 words, as it is plausible to assume that a two year old child has been exposed only to a subset of the words of a language. The value of the parameter $b$ can be estimated from the Zipf plot of the language, as it roughly corresponds to the average size of the groups of words with the same frequency but different ranks. Form the data examined it turns out that the choice $b \approx 4$ is rather plausible. Thus, the parameter γ is the only parameter that is calculated through a non-linear fitting procedure. We have arbitrarily chosen to set $\varepsilon$=200, but as one can easily see it is the ratio ε/γ that appears in equations 8-9.

*e. Effort needed for communication*

A rather simple, yet important, result can be derived from the model by calculating the derivative of the energy with respect to the temperature $C = \frac{\partial E}{\partial T}$, which is a measure of how much effort is needed in order to increase communication willingness (i.e. the temperature) by one degree.

---

[2] In order to derive Eq 8 it is useful to imagine that the most frequent word (value $k$=1) corresponds to the 'null' word.
[3] The results are not sensitive in the particular choice of the threshold as long as it is small enough. Choosing a large threshold physically means that we do not observe the children for adequate time, and thus we cannot record words that are rarely used.

It turns out that by differentiation of $E$ we get:

$$C = \varepsilon \left[ \frac{k_B \varepsilon}{(k_B T - \varepsilon)^2} - \left( \frac{b N \varepsilon (\ln(\frac{b}{N}))^2 (\frac{b}{N})^{\varepsilon/k_B T}}{k_B T^2 (b - N(\frac{b}{N})^{\varepsilon/k_B T})^2} \right) \right] \qquad (10)$$

In Figure 3 we present a plot of this quantity assuming that the number of speakers is constant and rather large so that $\varepsilon$ is constant. We use an arbitrary system of units setting $\varepsilon=200$ and $k_B=1$. We conclude that, although it is not so difficult to make a group of people that does not communicate at all ($T=0$) to start communicating, this task becomes more and more difficult if the group has already some small degree of communication willingness. There is, however, a maximum value for $C$ at a certain temperature and when we go over this temperature, a communication task is becoming easier.

*f. Bilingual Children Learning Curves*

We now use this model (Equation 8) in order to study the vocabulary size of bilingual children. In the case of bilingual children we expect that $N$ will be greater than in monolingual children, but also b will be greater as there will be more "basic" words. This is a prediction of the model that can be formally tested (Figure 4). It seems to disagree with the dogmatic belief that "bilingual children often talk late" and agree with the currently acceptable view that concerning verbal abilities it makes no difference if children grow in monolingual or bilingual environment [19].

*g. Schizophrenia*

Can a mental illness be related to the present framework? It turns out that the exponent β for linguistic data of people suffering from schizophrenia is lower than the equivalent data for normal people [12]. From our point of view, (since $\beta=1/\alpha +1$, the exponent *a* is higher in schizophrenia) this implies that the temperature T for subjects with schizophrenia is lower than "normal" subjects, indicating the lower communication ability associated with mental illness.

*h. Coding and non-coding DNA*

It is well known, and rather intriguing that in higher organisms only a small fraction of the DNA sequence is used for coding proteins [20]. This fraction is called "coding DNA" in contrast to the remaining part which is called "non-coding DNA". In a rather pioneering work by Mantegna et al [20] it was investigated if there is a resemblance between the DNA sequence and language. The DNA sequence was partitioned in sections of fixed length and these sections were considered as words. "Frequency–rank" plots were constructed for both coding and non-coding DNA. Power-law behaviour was indeed observed. Our model implies that a low slope *a* of such a graph is associated with "high" temperature and thus with more effective communication. Intuitively, we would expect that the coding DNA is associated with more effective communication compared to the non coding DNA. Indeed in all cases studied in [20] the exponent *a* is found much higher (up to a factor of 2.2) for the "non coding DNA" compared to the "coding DNA". Is this just a coincidence or a signal that nature has found once more the optimal way in order to convey information?

**4. Discussion and Conclusions**

The model proposed here predicts that if we count the word frequency distribution of young children dividing them in age groups, then the resulting Zipf plots for each age group will also exhibit power law behaviour, but the slope will be different than that for adults. Some preliminary work in this direction using CHILDES[4] [24] data base, seems to agree with this prediction but due to the very small sample and the noise of the data this is hardly a convincing indication at this stage. An interesting quantity to measure is also the word frequency distribution in oral conversation rather than in written documents. We expect that oral communication will also follow the Zipf law, but the question is whether oral communication has the same "temperature" as the written one or not. Most likely the Zipf exponent will be different in these cases. This task, however, requires considerable recourses and interdisciplinary effort and is, thus, left for future work.

It would, however, be useful to compare the proposed model to other models published in the literature. The pioneering work of *Nowak et al* [1,17,21] is based on evolutionary game theory, so it is mainly concerned with the description of language at large time scales. The evolution of language is a task taking several generations to be accomplished. Our approach relies on equilibrium statistical mechanics. As, very vividly, Feynman describes: "When all *fast things* have happened and all *slow things* have not, then the system is in equilibrium" [22]. In language there are *fast* processes, such as the physicochemical changes in the brain when a word is memorized and *slow* processes such as language evolution or the genetic evolution of the speakers. We study here the regime between the two, so the time scale is different from that of *Nowak et al*. Another model proposed by *Abrams and Strogatz* [2] deals with language competition. This is still a rather *slow* process in a different area complementary to ours, as our model does dot deal with competition between different languages.

---

[4] The Greek child data in this corpus were donated by Ursula Stephany [25,26]. They were collected between 1971 and 1974 in natural speech situations in the homes and Kindergarten from four monolingual Greek children growing up in Athens, Greece.

The success of statistical physics in the description of large scale materials relies on the fact that quantum mechanics can be used in order to determine the energy states of a system from first principles. For language we do not have yet such a theoretical description. We have to guess the form of the "energy states" so that it does not disagree with measured probability distributions such as the Zipf law. Is there a way to theoretically predict the form of the Hamiltonian? This is exactly the same as to ask whether Zipf law can be derived from first principles, as this derivation provides immediately a way to determine the word's energy states from first principles. Thus, for example, the pioneering work of Cancho and Sole [23] who attempt to derive Zipf law using the principle of least action could be related to our approach, much like the way that quantum mechanics and statistical mechanics are related. The conceptual importance of previous studies [1, 2, 17] is that they propose methods based on evolutionary game theory or differential equation models, thus bridging the gap between linguistics and mathematics or linguistics and evolutionary biology. Our treatment indicates that also the gap between linguistics and physics is not as large as it originally seems.

In summary, we have studied human language using the framework of equilibrium statistical mechanics. Within this framework we are able to qualitatively explain basic properties of language, such as the universality of the exponent of Zipf law and to predict reasonably well the form of the vocabulary size versus age curve of young children. The results obtained here are not particularly sensitive in the chosen values of the parameters and remain qualitatively the same for a wide range of different values.

It seems that statistical physics provide an interesting formulation for the study of language and can unify aspects, like the frequency distribution of words and children's vocabulary learning rate, which initially seem completely different.

We hope that the present paper will motivate new interdisciplinary work which will allow the verification or rejection of the model. This is essential, since some of the linguistic characteristics of interest are notoriously difficult to count and require combined forces of several groups.

**Acknowledgements**

This work was partially supported by the Greek Ministry of Education via Herakleitos and Pythagoras Projects.

**FIGURES**

1. Zipf plot for the word frequency in Greek language. Data taken from the corpus described in Ref[16]. Points are the experimental data and straight line is the best fit. The slope is -0.96. Notice the existence of groups of words with the same frequency and different rank.

2. Vocabulary size as a function of child age. The solid line is the prediction of our model. The points are averages of experimental data taken from Huttenlocher Ref [18]. The parameter values used were $\varepsilon=200$, $k_B=1$, $N=6000$ words and $b=4$. The values of the constant determined by non- linear fitting were $\gamma =5.2$

3. "Heat Capacity" $C$ vs temperature for a large group of people. The parameter values used were $\varepsilon=200$, $k_B=1$, $N=6000$ words and $b=4$. The quantity $C$ is indicative of the effort needed to increase the communication willingness of a group of people.

4. Comparative predictions for the vocabulary size between bilingual (upper curve) and monolingual (lower curve) children. The parameter values used were $\varepsilon=200$ and $k_B=1$ is used. The value $\gamma =5.2$ was used for both curves. For the monolingual curve we have used $N=6000$ words and $b=4$. For the bilingual curve we have used $N=12000$ words and $b=8$.

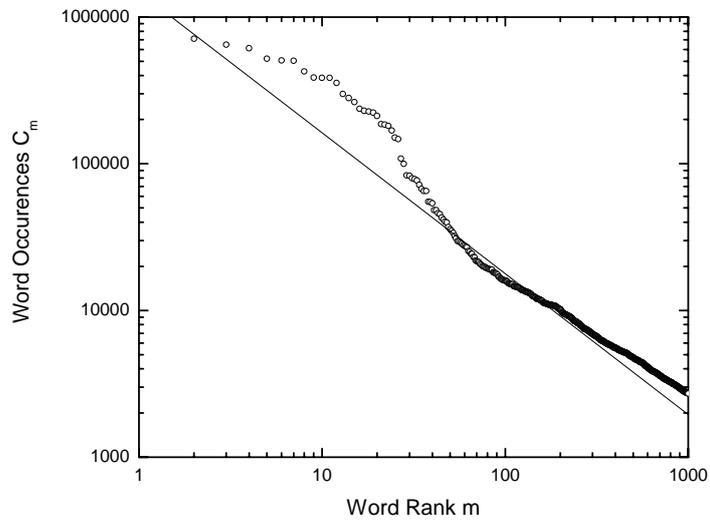

Figure 1

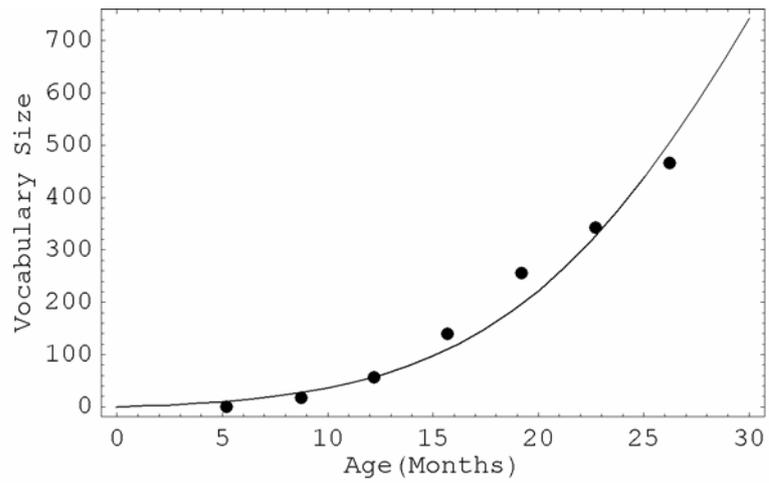

Figure 2.

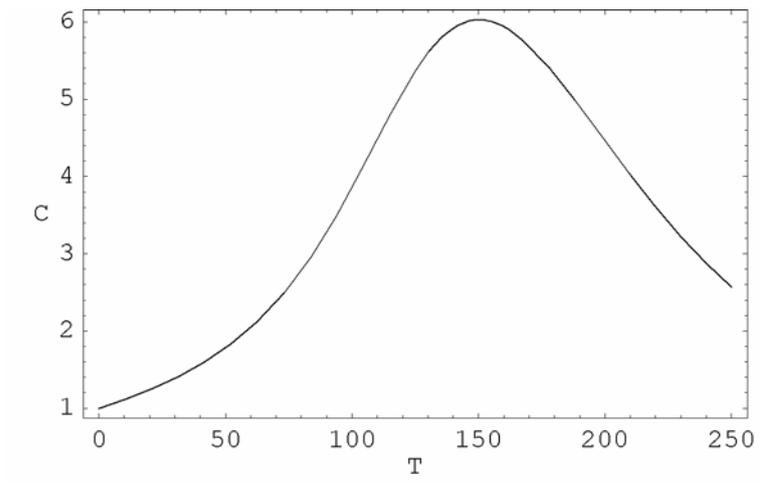

Figure 3

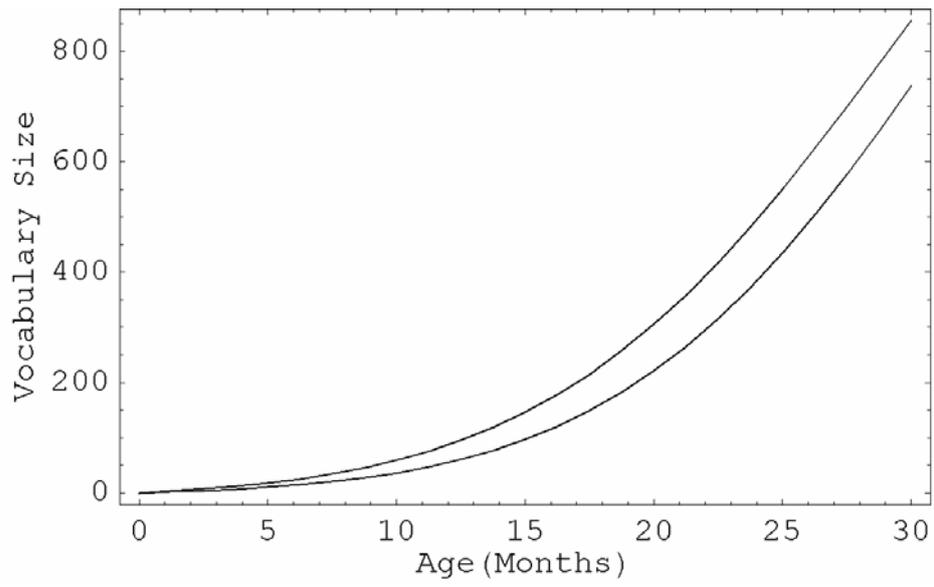

Figure 4